\documentstyle[12pt]{article}
\def\hybrid{\topmargin 0pt      \oddsidemargin 0pt
	\headheight 0pt \headsep 0pt
	\textheight 9in         
	\textwidth 6.25in       
	\marginparwidth .875in
	\parskip 5pt plus 1pt   \jot = 1.5ex}

\catcode`\@=11
\def\marginnote#1{}
\newcount\hour
\newcount\minute
\newtoks\amorpm
\hour=\time\divide\hour by60
\minute=\time{\multiply\hour by60 \global\advance\minute by-\hour}
\edef\standardtime{{\ifnum\hour<12 \global\amorpm={am}%
	\else\global\amorpm={pm}\advance\hour by-12 \fi
	\ifnum\hour=0 \hour=12 \fi
	\number\hour:\ifnum\minute<10 0\fi\number\minute\the\amorpm}}
\edef\militarytime{\number\hour:\ifnum\minute<10 0\fi\number\minute}

\def\draftlabel#1{{\@bsphack\if@filesw {\let\thepage\relax
   \xdef\@gtempa{\write\@auxout{\string
      \newlabel{#1}{{\@currentlabel}{\thepage}}}}}\@gtempa
   \if@nobreak \ifvmode\nobreak\fi\fi\fi\@esphack}
	\gdef\@eqnlabel{#1}}
\def\@eqnlabel{}
\def\@vacuum{}
\def\draftmarginnote#1{\marginpar{\raggedright\scriptsize\tt#1}}

\def\draft{\oddsidemargin -.5truein
	\def\@oddfoot{\sl preliminary draft \hfil
	\rm\thepage\hfil\sl\today\quad\militarytime}
	\let\@evenfoot\@oddfoot \overfullrule 3pt
	\let\label=\draftlabel
	\let\marginnote=\draftmarginnote
   \def\@eqnnum{(\theequation)\rlap{\kern\marginparsep\tt\@eqnlabel}%
\global\let\@eqnlabel\@vacuum}  }

\def\numberbysection{\@addtoreset{equation}{section}
	\def\theequation{\thesection.\arabic{equation}}}

\def\underline#1{\relax\ifmmode\@@underline#1\else
	$\@@underline{\hbox{#1}}$\relax\fi}

\def\titlepage{\@restonecolfalse\if@twocolumn\@restonecoltrue\onecolumn
     \else \newpage \fi \thispagestyle{empty}\c@page\z@
	\def\thefootnote{\fnsymbol{footnote}} }

\def\endtitlepage{\if@restonecol\twocolumn \else  \fi
	\def\thefootnote{\arabic{footnote}}
	\setcounter{footnote}{0}}  
\catcode`@=12
\relax

\def\demi{{1\over 2}}
\def\quart{{1\over 4}}

\def\d{\delta}

\def\ee{\eea}
\def\be{\bea}

\def\tr{\mathop{\rm tr}}
 \def\Tr{\mathop{\rm Tr}}
\def\beq{\begin{equation}}
\def\eeq{\end{equation}}
\def\bea{\begin{eqnarray}}
 \def\eea{\end{eqnarray}}

\def\bar{\overline}
 
 \def\nn{\nonumber}

\def\d{{\cal D}}
\def\I{{\cal I}}
\def\t{{\theta}}

\def\ic{\int_0^T}

\def\vq{\vec q}

\def\P{\Psi}
\def\bP{\bar \Psi}

\def\vP{\vec \P}
\def\vbP{\vec {\bP}}

\def\l{\lambda}
\def\vl{\vec \l}

\def\eij{\epsilon^{ij}}

\def\V{{\dot q}_i+{{\delta V}\over{\delta q_i}}}
\def\VP{{\dot \P}_i+{{\delta ^2 V}\over{\delta q_i \delta q_j}}\P_j}

\def\et{\epsilon(t)}

\def\v{{\dot q}_i+f{{\eij q_j}\over{\vec q^2}}}
\def\VPt{{\dot \P}_i+f{{\delta ^2 \t}\over{\delta q_i \delta q_j}}\P_j}
\def\p{\vec p}
\def\bQ{\bar Q}
\def\de{\delta}
 \def\al{\alpha}

 \def\demi{{1\over 2}}
\def\quart{{1\over 4}}

\relax
\numberbysection
\hybrid
\begin{document}

\begin{titlepage}
 \hfill Racah Institute 151-Paris LPTHE 93-28
\vskip 38 truemm
\begin{center}

{\large\bf On the Calculability of Observables in Topological Quantum
Mechanical Models }
\\[1in]
        {\bf    Laurent Baulieu}\footnote
{{	Tour 16,
	4 place Jussieu,
 75005 Paris, France}\hfill email address: Baulieu@lpthe.jussieu.fr
  }\\
 LPTHE, Universit\'e Pierre et Marie Curie, Paris\\

and\\
{\bf    Eliezer Rabinovici}\footnote{email address: Eliezer@vms.huji.ac.il}
\footnote{Work supported by the Israeli Accademy of Sciences and the
Bi-National
American-Israeli Science Foundation}\\
    	Racah Institute of Physics,  Hebrew University, Jerusalem\\

\end{center}

\vskip 1cm

\begin{quotation}
{\bf Abstract }
We consider a superconformal quantum mechanical system which has
been chosen on the basis of a local BRST topological invariance.
We suggest that it  truly leads to topological observables
which we compute. The absences of a ground state and of a mass gap are
special features of this system.    \end{quotation}
 \end{titlepage}

\newpage

\section{Introduction} During the last years, Topological Quantum Field
Theories
have emerged as possible realizations of general coordinates invariant
symmetries \cite{wittendona}\cite{review}. These theories produce
space-time metric independent correlations functions. For ordinary
gauge theories, the well known phases are the massive Higgs phases and
Coulomb and confinment phases. One of the great achievements of their
study was the understanding that these phases have universality
properties in the sense that their ulttraviolet behaviour are the same
and that the short distance behaviour of the massive phase is milder
than what one would expect for massive vector fields. It would be
rewarding if Topological Quantum Field Theories, which have striking
finitness properties at short distance, could be identified as a given
phase of systems with an invariance under general reparametrization,
so that the other phases, like quantum gravity, would also reveal
finitess properties.

In ordinary gauge theories the study of the behaviour of order parameters and
relevant operators has been a usefull tool to survey the possible phases. The
interesting question of finding the relevant order parameters for general
ccordinate invariant systems is yet unresolved.

In Topological Quantum Field Theories, an important operator which is
at disposal is the BRST operator $Q$, such that the Hamiltonian is
$H=\demi[Q,\bar Q]$. One usually defines the physical Hilbert space as
the cohomolgy of $Q$ (states which are annihilated by $Q$ without
being the $Q$ transformation of other states). This definition of the
physical Hilbert space is perfectly suited for ordinary gauge
theories. For Topological Field Theories there are doubts on the
validity of this definition and their could be other relevant
observables than those defined by the BRST cohomology. One may think
that the failure of finding states annihilated by $Q$ but the $Q$
exact ones would signal the transition to the topological phase. In
supersymmetric theories a similar situation indeed implies the
breaking of supersymmetry and would make possible the relevance of $Q$
exact observables.  Actually the absence of states annihilated by the
constraints is sometimes the symptom of a global gauge anomaly. It may
also occur that a more refined form ofthe BRST symmetry should be
introduced, the cohomology of which would contain the $Q$ exact
observables.

In the Topological Quantum Field Theories such as those originally
introduced in \cite{wittendona} which can be understood as the gauge
fixing of topological invariants \cite{bs}, the observables can be
seen to be $Q$ exact. In the other topological models based on first
order actions like the Chern-Simons action \cite{cs}, formal argument
show that the situation is similar. In all these models one sees
furthermore that, in addition to the global topological BRST symmetry,
a local version of it seems to characterize the supersymmetric
potential \cite{ab}\cite{rak}. However, little is known about how the
supersymmetry and ghost number conservation laws are truly broken to
yield non vanishing mean values of topological observables, and also
how one computes these mean values.

In this note we wish to study these questions in the framework of a
simplest model which shares the characteristics of the more
sophisticated models based on the gauge fixing of topological
invariants.  This model is a supersymmetric quantum mechanical system
defined on a punctured plane. It has been chosen under the condition
that it has an additional quasi-local symmetry.  This symmetry turns
out to be quite similar to a conformal supersymmetry and thus we can
use technics already introduced in \cite{fubini}. It is known that by
a Wick rotation any supersymmetric quantum mechanics action can be
understood as a topological action \cite{review} and moreover the
potential of our model is based on the criteria of a local BRST
invariance which we believe warrantees its non triviality
\cite{ab}. This model is  solvable, and we   verify that
due to the superconformal properties of its potential it has
some rather special features that
we study in details. The spectrum is continuous and contains all $E>0$
states but not the state with energy $0$ which is not even plane wave
normalizable. Thus supersymmetry is broken. Let us recall that Witten
suggested that non normalizability of the wave function could be a
general criteria for the signal of a phase transition
\cite{transition}. We then verify the existence of $Q$ exact
topological observable of which we compute the mean values.  Moreover,
we notice that this groundstateless system possesses an interesting
instanton structure and that all states are infinitely degenerate with
respect to the discrete angular momentum quantum number. It seems that
the answer that we obtain for topological observables means that in
the path integral one gets one single contribution from each each
instanton sector. The observable that we get is simply related to a
generator of a superconformal algebra. We have also computed the
Witten index and observed that what contributes to its value is the
non normalizable and measure zero part of the spectrum.  As far as the
exploration of the model is concerned we have adopted an open
attitude: we could have rejected it as having a global anomaly, but we
didn't and attempted to extract from it scale independant features.

\section{The Model}

We wish to work with a simple topological classical Lagrangian that is
a candidate to generate a topological quantum mechanics. We consider
as a target space a plane from which we exclude the origin, so that
one has a non trivial, although very simple topological structure
defined by the winding number around the origin of the trajectories of
a particle.  We denote the time by the real variable $t$ and the
Euclidian time by $\tau$, with $t=i\tau$ and $\tau$ real. The
cartesian coordinates on the plane are $q_i$, with $i=1,2$. We select
trajectories with periodic conditions, namely such that between the
initial and final times $t=0$ and $t=T$ the particle ends up at its
starting point so an integer value of the winding number can be
assigned to its trajectory.

 From our understanding of the nature of a topological field theory
\cite{bs}, we start from a topological classical action
$\I_{cl}[\vec{q}]$. $\I_{cl}[\vec{q}]$ must not depend on the time
metric. This condition is satisfied if it is the integral of a locally
closed form. The natural candidate is \be \I_{cl}[\vec{q}] = \int\
fd\t &=& \ic \ d\tau f\dot\t(\tau)\nn\\ &=& \ic \ d\tau\ f{{\eij \dot
q_i q_j}\over{\vec q^2}}\ee where $f$ is a real number.  This action
measures the winding number of the particle times $f/2\pi$. It shares
analogy with the second Chern class $\int d^4x \tr F\wedge F $ where
$F$ is the curvature of a Yang-Mills field. Here and in what follows
the symbol $\dot X $ denotes ${{d X }\over{d\tau}}$.

To obtain the Topological Quantum Theory associated to our space, we
need to give sense to the Euclidian path integral
\be \label{pa1}
\int \d [\vec q]\exp -\I_{cl}[\vec{q}]
\ee
as well as to compute topological quantities from Green functions
\be \label{pa6}
{\rm Topological} \   {\rm information}=\int \d [\vec q]{\it{O}}\exp
 -\I_{cl}[\vec{q}]  \ee
where ${\it{O}}$ is a well chosen composite operator.

The difficulty for realizing this objective is that our action is
different from that of conventional quantum mechanics where classical
degrees of freeedom exist at the classical level and quantum
fluctuations occur around the solutions of equations of motion. Here
the Lagrangian is locally a pure derivative, the Hamiltonian vanishes
and one has no equation of motion.  On the other hand, one observes
that the action $\I_{cl}[\vec{q}]$ is invariant under the gauge
symmetry
\be\label{topsym}
\vec{q(t)}\to \vec{q(t)}+\vec{\epsilon(t)}\ee
where $\et$ is any given local shift of the particle position $q(t)$
which does not change the winding number of the trajectory.  Using the
BRST technique it is then possible to define the path integrals
(\ref{pa1}) and (\ref{pa6}) by a conventional gauge fixing of the
action $\I_{cl}[\vec{q}]$.

The BRST transformation laws associated to the symmetry (\ref{topsym})
are of the simple form
\be
s\vq=\vP\quad s\vP=0\quad s\vbP=\vl\quad s\vl=0
\ee
The anticommuting
fields $\vP(t)$ and $\vbP(t)$ are the topological ghosts and
antighosts associated to the particle position $\vq(t)$.  $\vl(t)$ is
a Lagrange multiplier.  s acts on field functions as a differential
operator graded by the ghost number.

To get a gauge fixed action with a quadratic dependence on the
velocity $\vec{\dot q}$, one choses a gauge function of the type $\V$,
where the prepotential $V$ is an arbitrary given function of $\vq$.
This yields the following gauge fixed BRST invariant action $\I_{gf}$
which is supersymmetric
\be
\label{IGF} \I_{gf}
[\vq,\vP,\vbP,\vl]
&=&
\ic \ d\tau \left(
f \dot\t -s\bP_i(
\demi\l_i
-i\V)
\right)
\nn\\
&=&
\ic \ d\tau \left(f \dot\t -\demi\l_i^2
+i\l_i\left(\V\right)
-i\bP_i\left(\VP\right)
\right)
\ee

 The BRST symmetry $s\I_{gf}
[\vq,\vP,\vbP,\vl]=
0$ holds true independently  of the
choice of the function $V(\vq)$ and  the partition
function and the mean values of BRST invariant observables
\be
 Z=\int \d [\vq]\d [\vP]\d [\vbP]\d [\vl]\exp -\I_{gf}
\ee
\be
 <{\it{O}}>=\int \d [\vq]\d [\vP]\d [\vbP]\d [\vl] {\it{O}}\exp -\I_{gf}
\ee
are now well defined Euclidian path integrals. To understand $\V$ as a
gauge function for the quantum variable $\vq$, one may interpret the
result of the integration over the ghosts as a determinant.  The BRST
invariance of the field polynomial ${\it{O}}$ allows one to prove, at
least formally, the topological properties of $ <{\it{O}}>$. On the
other hand our knowledge of supersymmetric quantum mechanics tells us
that this mean value may depend on the class of the function $V$. What
happens is that in the case of topological field theories, the
Euclidian path integral explores the moduli space of the equation
$\V=0$, as a result of the gauge fixing.

The question of finding a symmetry principle which would select the
prepotential $ V(\vq)$ leading to interesting topological information
was investigated in \cite{ab}. The idea is to ask for the invariance
of the action under a symmetry which is more restrictive than the
topologocal BRST symmetry, namely a local version of it, for which the
parameter becomes an affine function of the time, with arbitrary
infinitesimal coefficients. One requires \be
\delta _l\I_{gf}[\vq,\vP,\vbP,\vl]=0
\ee
where the "local" BRST transformations $\delta _l$ are
\def\at{\eta(t)}
\be
\delta l\vq=\at\vP
\quad
\delta _l\vP=0
\quad
\delta _l\vbP=\at\vl-\dot\at\vq
\quad
\delta _l\vl=\dot\at\vP
\ee
and $\at=a+bt$ where $a$ and $b$ are constant anticommuting
parameters. The idea of local BRST symmetry was considered in
\cite{STORA} for the sake of interpreting higher order cocycles which
occurs when solving the anomaly consistency conditions, and has been
shown to play a role in topological field theories in \cite{rak}.

Imposing this local symmetry implies that $V$ satisfies the constraint
\cite{ab} \be
{\delta V\over \delta q_i}+q_j\ {\delta\sp 2 V\over \delta q_i\ \de
q_j}=0
\ee
This constraint is solved for $V(\vq)=f\t$ where $\t$ is the angle
such that $q_1+iq_2=|\vq|\exp i\t$ and f is a number
\cite{ab}\footnote{For the case of one variable $x$ we would obtain
$V=\log x$, with quite similar properties of the supersymmetric
system, but the geometrical interpretation would be less clear and no
meaningful observable exists}. By putting this value of $\V$ in
(\ref{IGF}) and eliminating the Lagrange multiplier $\l$ by its
equation of motion we obtain \be \I_{gf} [\vq,\vP,\vbP]=
\ic \ d\tau \left( \demi \dot q_i^2+{f^2\over{2\vq^2}}
-\bP_i\left(\VPt\right)
\right)
\ee
Notice that
\be
{{\delta ^2 \t}\over{\delta q_i \delta q_j}} &=&
{1\over{\vq^2}}
\pmatrix{
-\sin 2\t&
\cos 2\t
\cr
\cos 2\t&
\sin 2\t
}_{ij}
\nn\\
&=&
{1\over{\vq^2}}\pmatrix{
 \cos  \t&
-\sin\t
\cr
\sin\t&
\cos\t
}
\pmatrix{
0 &
-1
\cr
-1&
0
}
\pmatrix{
 \cos  \t&
\sin\t
\cr
-\sin\t&
\cos\t
}_{ij}
\ee
The superconformal potential $1/\vq^2$ has been already studied in
\cite{fubini}\cite{rafu}. In the next section we shall compute the
observables which seems interesting to us from the topological point
of view in the canonical quantization formalism.  We will show that a
very specific supersymmetry breaking mechanism occurs and implies the
existence of non vanishing $Q$ exact observables which are metric
independent as well as of a fractional Witten indexD.

We believe that the signal that the theory truly carries some
topological information is the existence of an interesting instanton
structure. Let us remember that, from our gauge fixing in the
Euclidian time region, we have obtained an action whose bosonic part
is the square of the gauge function. It follows that the solutions to
the Euclidien equations of motion can be written as
\def\v{{\dot q}_i\pm f{{\eij q_j}\over{\vec q^2}}}
\def\VPt{{\dot \P}_i\pm f{{\delta ^2 \t}\over{\delta q_i\delta q_j}}\P_j}
\be \v=0
\ee
\be
\VPt=0
\ee
If we introduce
$z=q_1+iq_2$ and $\P_z=\P_1+i\P_2$, with $sz=\P_z$, we can write
these equations as
\be
\dot z+{i\over{z^*}}=0
\ee
\be
\dot \P_z-{i\over{(z^*)^2}}\P_z^*=0
\ee
Assuming  periodic boundary conditions,
 the solutions for $\vq$ are circles described at
constant velocities and indexed by an integer $n$
\be
z^{(n)}=\sqrt{T\over{2n\pi}}\exp -i{{2nt}\over T} \quad\quad n\in Z
\ee
while for the ghost
\be
\P_z^{(n)}=\eta \exp i {{2nt}\over T}
\ee
where $\eta$ is a constant fermion. The Euclidian energy and angular
momentum of the action evalueted for these field configurations vanish
for all valuese of $n$.

Due to the existence of these degenerate zero modes of the action we
expect that BRST invariant observables should exist and that their
mean values should be non zero as well as energy and time
reparametrization independent. The corresponding numbers should to be
expressable as a series over an integer related to the one which label
the instanton solutions. This is the conjecture that we shall verify
in the next section.

\section {Hamiltonian Quantization}

When possible, it is worthwhile to compute observables in the canonical
formalism. We  do a Wick rotation to recover the real Minkowski time $t$ by
setting  $\tau= it$, and change the quantum mechanichal variables  into
operators.  The   Hamiltonian associated to the action $\I_{gf}$ is
 \be H=\demi\p^2
+{f^2\over{2\vq^2}}
-f\bP_i{{\delta ^2 \t}\over{\delta q_i \delta q_j}}\P_j
\ee
where  the quantization rules are
(  remember that $q_i=(x,y)$ stands for the
  cartesian coordinates on the plane)
\be
[p_i,q_j] =-i\delta_{ij}
\quad
\quad
[\bP_i,\P_j]_+=\delta_{ij}\nn
\ee
\be
[\P_i,\P_j]_+=[\bP_i,\bP_j]_+=
[\Psi_i,p_j]=[\bP_i,p_j]=[\Psi_i,q_j]=[\bP_i,q_j]=0\ee
By construction $H$  can be
written as
 \be
H=\demi\left\{ Q,\bQ
\right\}
\ee with
\be
Q=\P_i(p_i+if{{\delta\t}\over{\delta q_i}}
)\quad\quad
\bQ=\P_i(p_i-if{{\delta\t}\over{\delta q_i}}
)\ee
Following \cite{rafu}, we use the following matricial representation for the
ghost and antighost  operator
\be
\P_1=
\pmatrix{
0&
1&
0&
0
\cr
0&
0&
0&
0
\cr
0&
0&
0&
1
\cr
0&
0&
0&
0  }
\quad\quad
\P_2=
\pmatrix{
0&
0&
-1&
0
\cr
0&
0&
0&
1
\cr
0&
0&
0&
0\cr
0&
0&
0&
0  }
\ee
 One has  $\bP=\P^\dagger$ and
$p_i=-i\delta/\delta q_i$.
In  this representation \be
H=
 \pmatrix{
H_0&
0&
0&
0
\cr
0&
H_{11}&
H_{12}&
0
\cr
0&
H_{21}&
H_{22}&
0
\cr
0&
0&
0&
H_2  }
\ee
where
\be
H_0=H_2=-{1\over 2r}{\de\over \de r}
r{\de\over \de r}-{1\over 2r^2} {\de^2\over \de\t^2}
+{f^2\over 2r^2}
\ee
and
\be
\label{Hone} \pmatrix{
 H_{11}&
H_{12}
\cr
H_{21}&
H_{22} }
&=&H_0\delta_{ij}+
f{{\delta ^2 \t}\over{\delta q_i \delta q_j}}
\nn\\
&=&
R_{-\t}
 (
-{1\over 2r}{\de\over \de r}
r{\de\over \de r}+  {{f^2+1-{\de^2\over \de\t^2} }\over {2r^2}}
-{1\over  r^2}
\pmatrix{
0&
f+i {\de \over \de\t }
\cr
f- i{\de \over \de\t } \t&
 0 }
 )
R_\t
\nn\\
\ee
where $r$ and $\t$ are the polar coordinates on the plane and
\be
R_\t=
\pmatrix{
\cos\t&
\sin \t
\cr
-\sin \t&
 \cos\t }
\ee
 The spectrum of $H$ is straightforward to derive in this
representation.  One uses the usual strategy based on the fact that if
an eigenstate of $H$ has energy $E$, its $Q$ and $\bQ$ transforms are
either zero or an eigenstate of $H$ with the same energy. States are
labelled by their non negative energy $E$, angular momentum n and
fermion number $\alpha$, that is ghost number. We denote them as
$|E,n,\alpha>$. For each value $E$ and $n$, one has four states
labelled by $\alpha=1,2,3,4$.  The states with $\al=1$ and $\al=4$ are
respectively anihilated by $Q$ and $\bQ$. This is due to the fact that
states $|\phi >$ which are BRST invariant, $ Q |\phi >=0$, are such
that
\be
|\phi >=\pmatrix{
| {E,n}>
\cr
0
\cr
0
\cr
0 }
\quad \quad
\bQ|\phi>=
\pmatrix{
 0
\cr
(p_1-if{{\delta\t}\over{\delta q_1}})| {E,n}>
\cr
- (p_2-if{{\delta\t}\over{q_2}})| {E,n}>
\cr
0 }
\nn\ee
\be
H|\phi >=
\pmatrix{
H_0| {E,n}>
\cr
0
\cr
0
\cr
0 }
\ee
One has similar relations for states $|\bar\phi >$ satisfying  $\bQ|\bar\phi
>=0$.

Let us  define $g_{E,n}=<r,\t|E,n>$. This function is the solution of the
equation \be\label{bessel1}
<r,\t|H_0  |E,n>=
\left(
-{1\over 2r}{\de\over \de r}
r{\de\over \de r}-{1\over 2r^2} {\de^2\over \de\t^2}
+{f^2\over 2r^2}
\right)
g_{E,n}=
Eg_{E,n}
 \ee
$g_{E,n}$ is also the solution of the ghost number $2$ equation
$<r,\t|H_2|E,n>=E|E,n>$.
Its knowledge is
sufficient to get the full  spectrum for $E\neq 0$. One has indeed  \be
| {E,n,1}>
=\pmatrix{
| {E,n}>
\cr
0
\cr
0
\cr
0 }
\quad\quad
| {E,n,2}>
={1\over\sqrt{E}}\bQ | {E,n,1}>
\nn\\
| {E,n,4}>
=\pmatrix{
0\cr
0
\cr
0
\cr
| {E,n}>
 }
\quad\quad
| {E,n,3}>
={1\over\sqrt{E}}\bQ | {E,n,4}>
\ee
The diagonalization of the part with ghost number one of the
Hamiltonian (\ref{Hone}) amounts to solve the equations
\be
\label {bessel2}
\left(
-{1\over 2r}{\de\over \de r}
r{\de\over \de r}+{f^2+1-
{\de^2\over \de \t^2}\pm 2\sqrt{ f^2-{\de^2\over \de \t^2}} \over {2r^2}}
\right)
g_{E,n,\pm}=
Eg_{E,n,\pm}
 \ee
which are of the same type as  (\ref{bessel1}).

 To solve (\ref{bessel1}) and (\ref{bessel2}) we set
\be
g_{E,n}&=&{1\over\sqrt{2\pi}}\exp in\t \ f_{E,n}(r)\quad\quad n\in Z\nn\\
g_{E,n,\pm}&=&{1\over\sqrt{2\pi}}\exp in\t \ f_{E,n,\pm}(r)\quad\quad n\in Z
\ee
For $E\neq 0 $, $f_{E,n}(r)$ and  $f_{E,n,\pm}$ are expressable as a Bessel
function $J_\nu( {\sqrt 2E }r)$ of order $\nu$, with
\be
\label{sol}
f_{E,n}(r)={1\over\sqrt{2}}J_{\sqrt{n^2+f^2}}( {\sqrt 2E }r)
\ee
and
\be
\label{sol1}
f_{E,n,\pm}(r)={1\over\sqrt{2}}\exp in\t J_{\sqrt{
f^2+1+n^2\pm2\sqrt{ f^2+n^2}
}}( {\sqrt 2E }r)
\ee

These states are normalizable as plane waves in one dimension. This is
a consequence of the continuity of the spectrum in the radial
direction. They build an appropriate basis of stationary solutions
since, with the normalization factor which is explicit in (\ref{sol}),
one has $\sum_n\int _{E>0}dE |E,n><E,n|=1$.  On the other hand, for
$E=0$, the Schr\"odinger equations (\ref{bessel1}) and (\ref{bessel2})
have no admissible normalizable solution.  Thus we have a continuum
spectrum, bounded from below, with a spin degeneracy equal to $4$ and
an infinite degeneracy in the angular momentum quantum number $n$. The
peculiarity of this spectrum is that there is no ground state, since
we have states with energy as little as we want, but we cannot have
$E=0$. This is a consequence of the conformal property of the
potential $1\over{|\vq|^2}$.

Since we cannot reach the energy zero which woud be the only $Q$ and
$\bar Q$ invariant state, we conclude that supersymmetry is broken.

It is useful for what follows to redefine the ghost and antighost
operators into
\be
 \pmatrix{
\P_r
\cr
\P_\t }
= \pmatrix{
 \cos\t&
\sin \t
\cr
-\sin \t&
 \cos\t }
\pmatrix{
\P_1
\cr
\P_2 }
\quad
\pmatrix{
\bP_r
\cr
\bP_\t }
= \pmatrix{
 \cos\t&
\sin \t
\cr
-\sin \t&
 \cos\t }
\pmatrix{
\bP_1
\cr
\bP_2 }
\ee
These rotated ghost operators satisfy similar anticommutation
rotations as the $\P_i$ and $\bP_i$. On the other hand, notice that
\be
[
{\delta\over{\delta \t}},\ \pmatrix{
\bP_r
\cr
\bP_\t }]_+=
  \pmatrix{
 \bP_\t
\cr
-\bP_r }
\quad
\quad
 [
{\delta\over{\delta \t}},\ \pmatrix{
\P_r
\cr
\P_\t }] =
 \pmatrix{
 \P_\t
\cr
- \P_r }\ee
\be
[
{\delta\over{\delta r}},\ \pmatrix{
\bP_r
\cr
\bP_\t }] =
 [
{\delta\over{\delta r}},\ \pmatrix{
\P_r
\cr
\P_\t }]_+=0\ee
One has the following expression of $Q$ and $\bQ$ which will be used
in the next section
\be Q=-i\P_r{\delta\over{\delta r}}
-i{1\over{  r}}
\P_\t({\delta\over{\delta \t}}-f)
\quad\quad
\bQ=-i\bP_r{\delta\over{\delta r}}
-i{1\over{  r}}\bP_\t({\delta\over{\delta \t}}+f)
\ee
These expressions in curved coordinates could be obtained from the
general formalism of \cite{macfarlane}].

 \section {Computation of BRST Invariant Observables} In the last
section we have seen that supersymmetry is broken in a very special
way. This opens the possibility of having non vanishing BRST-exact
Green functions which are topological in the sense that they are scale
independent, that is independent of time, or energy, rescalings.

{}From dimensional arguments the candidates for     such
commutators are  \be
{\it{O}}_\t=[Q,r\bP_\t]_+ =[\bQ,r\P_\t]^\dagger_+
\quad\quad
{\it{O}}_r=[Q,r\bP_r]_+=[\bQ,r\P_r]^\dagger_+
\ee
The   mean values of these operators
between normalized states  are
\def\B{J_{\sqrt{n^2+f^2}}}
\be\label{obs}
{{<E,n| [Q,r\bP_\t]_+|E,n>}\over{<E,n|E,n>}}=n+if
\ee
and
\be\label{nobs}
{{<E,n| [Q,r\bP_r]_+|E,n>}\over{<E,n|E,n>}}=\lim_{L\to\infty}
{{L^2\B^2(L)}\over{\int_0^L dr\B(r)}}
\ee
The last quantity is bounded but ill-defined, so we reject it. We get
therefore that for any normalized state $|\phi_n>=\int dE
\rho(E)|E,n>$ with a given angular momentum $n$, the expectation value
of $[Q,r\bP_\t]_+ $ is
\be\label{obs4}
<\phi_n|[Q,r\bP_\t]_+|\phi_n>=n+if
\ee
indepently of the weighting function $\rho$.

If we now sum over all values of $n$, what remains is the topological
number
\be\label{obs1}
<{\it{O}}_\t>=\sum_n <\phi_n|[Q,r\bP_\t]_+|\phi_n>=
\sum_n n +if\sum_n1
\ee
 From a topological point of view, our result mean that there are two
observables, organized in a complex form, in the cohomology of the
punctured plane. The summation over the index $n$, that is the angular
momentum, could have expected from the formal argument that in the
path integral one gets a single finite contribution from each
instanton solution to the mean value of a topological observable, so
that
\be \label{pa2}
{\rm Topological} \ {\rm information}=\int \d [\vec q]{\it{O}}_f\exp
-\I_{cl}[\vec{q}]\sim \sum_n f(n)\ee Our computation shows the
existence of a BRST invariant observable with non zero mean value
which is{\it Q-closed}.  The supersymmetry breaking mechanism made
possible by our potential choice (on the basis of local BRST symmetry)
is responsible of this situation.  With other potentials than the one
that we have chosen , either supersymmetry would be unbroken, or a
mass gap would occur. In the previous case all Q-exact observable
would vanish; in the latter case they could be nonzero but they would
be scale dependent.

As another topological observable of the theory, we may consider the
Witten index
\cite{BSYM} \cite{Comtet}. The idea is that although there is no normalizable
vacuum in the theory, we can consider the trace
\be
\Delta=\Tr (-)^F\exp-\beta H
\ee
where the trace means a sum over angular momentum as well as over all
energy including energy zero, and $(-)^F$ is the ghost or fermion
number operator. The result should be finite because, although the
state with energy zero is not normalizable, it contributes only over a
domain of integration with zero measure. Indeed, since supersymmetric
compensations occur for $E\neq 0$ and provided one uses a BRST
symmetry preserving regularization, the full contribition to $\Delta$
should come from the domain of integration concentrated at $E\sim 0$,
while the topological nature of the theory should warranty that
$\Delta$ is non zero and independent on $\beta$.

By using the suitably normalized eigenfunctions of the Hamiltonian,
eqs.(\ref{sol}) and (\ref{sol1}), one can write the index $\Delta$ as
follows
\be
\Delta=\sum_n
\int_0^\infty
dE
&\exp& -\beta E
\int rdr  \demi (
 2\B^2(\sqrt{2E} r) \nn\\ &-&J_{\sqrt{ f^2+1+n^2+\sqrt{
f^2+n^2}}}^2(\sqrt{2E} r) -J_{\sqrt{ f^2+1+n^2-\sqrt{
f^2+n^2}}}^2(\sqrt{2E} r) )\nn\\\ee To compute this double integral
one needs a regularisation. Following for instance \cite{Comtet}, we
can use a dimensional regularization . Thus we change $dr$ into
$r^{\epsilon}dr$.  Then, the analytic comtinuation of the result when
$\epsilon\to 0$ is
\def\an{{\alpha_n}}
\be
\Delta
=\sum_n\demi\left(2\sqrt{ f^2+n^2}
-\sqrt{
f^2+1+n^2+\sqrt{ f^2+n^2}}
-\sqrt{
f^2+1+n^2-\sqrt{ f^2+n^2}}\right)\nn\\
\ee
As announced this result is independent on $\beta$. As a series, it
diverges logarimically as $\sum 1/n$ which is presumably the
consequence of the conformal invariance of the potential. We see that
the contribution of each topological sector is n dependant.

\section{Underlying superconformal properties}

We now explain our result of getting energy independant expectation
value of a Q-exact operator as a consequence of the underlying
superconformal structure of our model. Firstly, let us investigate the
possible consequences of having an operator $A$ which commutes with
the Hamiltonian $H$ and which is BRST exact, that is $A=[Q,X]$. $A$
has zero expectation values between physical states, that is $Q$ and
$\bar Q$ invariant states. For non physical states, that is states of
given non zero energy E, each eigen-value of the operator $A$ is a
priori implicitely dependent on the value of $E$, since although $A$
and $H$ are simultaneously diagonalizable, their eigenvalues are
generally connected.  However, it may happen that this dependance is
eliminated if, for each value of $E$, all possible eigenvalues of $A$
are allowed. This is exactly the situation which occurs in our model
where the observable shown in (\ref{obs}) is of the type of such an
operator $A$ and where one can verify that for all values of $E$ all
possible values of the spin and the angular momentum are allowed for.
We can can therefore consider the summation shown in (\ref{obs1}), or
the limit of (\ref{obs}) when $E\to 0$.

The existence of this observable could have been foreseen from the
superconformal properties of the potential. Consider first the case
where we would be in one dimension $x$, with the same potential in
$f/q^2$. In this case the candidate for a topological observable
doesnot lead to a well defined result. To prove this, following
\cite{fubini}, we define the following superconformal algebra
generators
\be
S= \Psi x
\quad\quad
\bar S=\bar \Psi x
\quad\quad
K=\demi x^2
\ee
One has the superconformal algebra relations
\be
\demi[\bar S, S]_+=K
\ee
\be
\demi[Q,\bar S]_+={f\over2}-\quart [\bar \Psi, \Psi]_+-iD
\ee
where $D$ is dilatation operator
\be
D=-\quart [x ,p]_+
\ee
One can verify that $D$ does not commute with $H$,
\be
lD,H]=-iH
\ee
which means that in this case the operator $A=\demi[Q,\bar S]_+$ is
not useful to obtain $E$ independant observables. What happen is that
$<D>$ leads to an undefinned result, similar to (\ref{nobs}), with $r$
replaced by the one dimensional variable $x$ and no well defined
observable exists in this situation.

Consider now  our case with   two dimensions. One defines
\be
S= \Psi_i x^i
\quad\quad
\bar S=\bar \Psi_i x^i
\ee
Due to the  the $O(2)$ invariance one also defines
\be
S'= \epsilon_{ij}\Psi_i x^j \quad\quad
\bar S'=\epsilon_{ij}\bar \Psi_i x^j
\ee
One can then verify that, independently of the value of the coupling f
\be
[Q,\bar S']_+=iJ\quad\quad [H,J]=0
\ee

It follows that the operator $J$, which is part of the superconformal
algebra of quantum mechanics in two dinensions, can be used as the
observable $A$ discussed at the beginning of this section. More
precisely, to interpret the obervables written in (\ref{obs4}), one
can one identify the operator $J$ and the eigenvalues $\sum n$, and
the identity operator with the eigenvalues $f\sum 1$.

We conclude this section by recalling that a way to regulate conformal
supersymmetric quantum mechanics and allow for the existence of a
ground state was proposed in \cite{fubini}. It was suggested to change
the time evolution generator, that is the Hamiltonian $H$ into
\be
H=\demi(aH+a^{-1}K)
\ee
As a result, the BRST operator is changed by terms of order $a^{-1}$,
a normalisable ground state occurs for which the mean values of our
observables vanish.  What is interesting is that the eigenstates of
the new time evolution operator, when expressed in terms of those of
$H$ build a spectrum not dissimilar to that involved in blackhole
physics \cite{fubini}.

 \section{discussion}

We have shown an example for which the requirement of local BRST
symmetry for topological quantum mechanics results in selecting a
superconformal quantum mechanichal system. As a result, the spectrum
of the theory has no ground state and a supersymmetry breaking
mechanism occurs, without the the presence of a dimensionful
parameter. Our goal was to understand the mechanism which provide
topological observables.  We observed that the special properties of
the potential allows the computation of energy independant quantities
although they are of mean values of BRST exact observables between non
zero energy states. These quantities deserve to be called topological
and they get a contribution from the whole spectrum of the theory. We
have also singled out the Witten index, in a computation which
includes a contribution from the non normalizable state of zero
energy. The generalization of these observations to quantum field
theory is an interesting open question.

We wish to thank T. Banks and S. Elitzur for very useful discussions.

\end{document}